\documentclass[a4paper,12pt]{article}

\usepackage[centertags]{ams math}
\usepackage{amssymb}

\usepackage{graphicx}
\usepackage{epsfig}
\usepackage{ulem}
\usepackage[english]{babel}
\usepackage{array}

\usepackage{latexsym}

\usepackage{yfonts}
\usepackage{mathrsfs}
\usepackage[mathcal]{euscript}

\pdfoutput=1
\usepackage{epsfig}
 \usepackage{jheppub}
\usepackage{mathdots}
\usepackage{MnSymbol}
\usepackage{multirow}
\newcommand{\nn}{\nonumber}
\newcommand{\be}{\begin{equation}}
\newcommand{\ee}{\end{equation}}
\newcommand{\bea}{\begin{eqnarray}}
\newcommand{\eea}{\end{eqnarray}}

\newcommand{\cI}{{\cal I}}

\newcommand{\cO}{{\mathcal O}}

\title{Bose Gas Modeling of  the  Schwarzschild Black Hole Thermodynamics}
\author{I.Ya. Aref'eva and I.V. Volovich}

\affiliation{Steklov Mathematical Institute, RAS}
\emailAdd{arefeva@mi-ras.ru}
\emailAdd{volovich@mi-ras.ru}

\abstract{
Black holes violate the third law of thermodynamics, and this gives rise to difficulties with the microscopic description of the entropy of black holes.
Recently, it has been shown that the microscopic description of the  Schwarzschild black hole  thermodynamics in $D = 4$ spacetime dimensions is provided by the analytical continuation of the entropy of Bose gas with non-relativistic one particle energy to $d =-4$ negative spatial dimension. 
In this paper, we show that  the $D=5$ and $D=6$
Schwarzschild black holes thermodynamics can be modeled  by the d-dimensional Bose gas, $d=1,2,3...$, with the one particle energy $\varepsilon(k)=k^\alpha$ under conditions $\alpha=-d/3$ and  $\alpha=-d/4$, respectively. In these cases the free energy of the Bose gas has  divergences and we introduce a cut-off and perform the minimal renormalizations. We also perform renormalizations
using analytical regularization and prove
that the minimal cut-off renormalization gives the same answer as the analytical regularization by the Riemann zeta-function. 
}

\begin{document}
\maketitle
\section{Introduction}
The  problem  with the microscopic  origin  of the Bekenstein-Hawking entropy \cite{BCH,Bekenstein:1973}
for the   Schwarzschild black holes  is that black holes do not satisfy the third law of thermodynamics  in its standard  formulation.
Therefore, such exotic thermodynamics behaviour of black hole cannot be obtained by using ordinary quantum statistical mechanics models which obey the third law, see discussion and refs in \cite{Arefeva:2023kpu}.\\

In \cite{Arefeva:2023kpu} we have shown that the entropy of the $D=4$  Schwarzschild black hole
\be
S_{BH}=\frac{\beta ^2}{16 \pi },  \qquad\beta=\frac{1}{T}\,,
\ee
where $T$ is the temperature, corresponds to the Bose gas in $d=-\,4$ {\it negative} spatial dimensions. This conclusion is obtained by using properties of the Riemann zeta function. The entropy
of the Bose gas in $d$-dimensional space  is proportional to
   \bea
 \label{Sdalpha-answer-i}
S_{BG}&\sim&  
   \left(\frac{d
   }{2}+1\right)
    \zeta \left(\frac{d
   }{2}+1\right)\,\beta^{-\frac{d}{2 }}\,,
   \eea
where  $\zeta$ is the Riemann zeta function. The expression \eqref{Sdalpha-answer-i} admits the analytical continuation for complex $d$, in particular for $d=-4$ we have
\bea
 \label{Sdalpha-answer-4}
S_{BG}&\sim&  \beta^2,
   \eea
  therefore, we get the entropy of the $D=4$ Schwarzschild black hole. 
  Note that the proportionality factor is a positive number and there is no divergences in this calculation.
\\

In this paper we show that some higher-dimensional black holes can be described using the Bose gas in positive dimensions. However, in these cases there are divergences that should be renormalized. 
We consider the  $d$ dimensional Bose gas with the kinetic term $k^{\alpha }$, in this case the free energy $F_{BG}$ is  proportional to
 \bea
 \label{fe-d-x}F_{BG}&\sim&
I(-\frac{d}{\alpha})\,\beta ^{-1-d/\alpha},\eea
where
\bea
\label{Is}
I(s)=\int _{0}^{\infty}\ln
\left(1-e^{- x}\right)\, \frac{dx}{x^{1+s}}.\eea

Of particular interest to us is the case with $d/\alpha=2-D$, since in this case we get
\bea 
 \label{fe-d-nx2}F_{BG}&\sim &I(D-2)\,\beta^{D-3}
,\eea
that coincides with  the Schwarzschild black hole dependence of the free energy on the inverse temperature $\beta$, $F_{BH}\sim \beta^{D-3}$.
However, the integral $I(s)$ diverges  for $s\geq 0$, and the formula \eqref{fe-d-x} has no immediate  meaning.
 To cure the formula \eqref{fe-d-x} we  introduce regularization in \eqref{Is}
and then perform renormalizations. We consider  two possible regularizations
 of the integral in \eqref{Is}: cut-off regularization and analytical regularization \cite{VSV}.
 In both cases we performed minimal subtractions and define $I_{ren}$ 
 and $\cI_{ren}$ in the first and second cases, respectively. We prove that both regularizations give the same answer, that explicitly means the validity of the identity \eqref{Eq} presented in Sect.\ref{Equiv}.
 In particular,  $D=5$ and $D=6$ black hole spacetime dimensions  correspond to the Bose gas model with $d/\alpha=-3$ and  $d/\alpha=-4$, respectively.
\\

The paper is organized as follows.
 In Sect.\ref{Setup} the Bose gas model with non-standard kinetic term is presented
 and two possible schemes of free energy renormalizations  are mentioned. 
 In Sect.\ref{CutR} the cut-off regularization is introduced and the minimal renormalization is performed. In Sect.\ref{AnalR} the analytical regularization is introduced and the its minimal renormalization is presented. 
 Sect.\ref{Equiv} the equivalence of the
 cut-off  minimal renormalization and minimal analytical renormalization is proved. In Sect.\ref{Examp} few explicit examples are presented and we conclude in Sect.\ref{Concl} with the discussion of obtained results.
 
\newpage
\section{Setup}\label{Setup}
We consider the Bose gas with
 kinetic term $\lambda (\vec k,\vec k)^{\alpha /2}$. 
 In d-dimensional case the free energy 
%integrating over spherical angles 
is \cite{LLV,VZ}

\bea
\label{fsdalpha-angle}F_{BG}&=&
\frac{\Omega_{d-1}}{ \beta}\left(\frac{L}{2\pi}\right)^d\int _{0}^{\infty}\ln
\left(1-e^{- \beta\,\lambda\,
 k^\alpha}\right)\,k^{d-1} dk,
 \eea
 where $\Omega_{d-1}=2\pi ^{d/2}/\Gamma(d/2)$
 and $\beta, \lambda,\alpha, L$ are positive constants,
 $d=1,2,3,...$. By changing the variable
 \be
 k=\left(\frac{x}{\beta \lambda}\right) ^{1/\alpha},\ee
 we get
 \bea
 \label{fe-d-1x}F_{BG}&=&
\frac{\Omega_{d-1}}{ \alpha\beta}\left(\frac{L}{2\pi}\right)^d\left(\frac{1}{\beta \lambda}\right)^{d/\alpha}
I(-\frac{d}{\alpha}),\eea
where
\bea
\label{I(s)}
I(s)&=&\int _{0}^{\infty}\ln
\left(1-e^{- x}\right)\, \frac{dx}{x^{1+s}}.\eea
For $d=1,2,3,...$ and $\alpha>0$ the integral in 
\eqref{I(s)}  converges
and 
\bea
I(s)=-\Gamma(-s)\zeta(1-s),\quad \Re s<-1.
\eea
However, as has been mentioned in Introduction,   the integral in \eqref{I(s)} diverges for  $s\geq 0$. To give a meaning for this formula for $s\geq 0$ we introduce regularizations. We consider two regularizations: cut-off regularization and analytical regularization.
We performed minimal subtractions and define $I_{ren}$ 
 and $\cI_{ren}$ in the first and second cases, respectively. Below we schematically describe both of them.

\begin{itemize}
\item Cut-off regularizations. In this case we start from
\bea\label{Isa}
I(s,a)&\equiv&\int _{a}^{\infty}\ln
\Big(1-e^{- x}\Big)\, \frac{dx}{x^{1+s}},\, a>0.\eea
We find a singular part of the asymptotics of  the integral $I(s,a)$ as $a\to 0$ in the form
\bea
S(s,a)=\sum_{i\geq 0} A_i\frac{\log a }{a^i}+\sum_{i\geq 1} C_i\frac{1 }{a^i}.\eea
Then we subtract this singular part
$S(s,a)$\bea
 I_{ren}(s,a)&=&
 I(s,a)-S(s,a),\eea
 and finally remove the regularisation
 \bea\label{Iren}
I_{ren}(s)&=&\lim _{a\to 0}I_{ren}(s,a).
\eea
\item  Analytical regularization.  In this case we start from the following representation 
\bea 
\label{repr-s} 
I(s)=\int _{0}^{\infty}\ln
\Big(1-e^{- x}\Big)\,\frac{dx}{x^{1+s}} =
 -\,\Gamma (-s)\,\zeta(-s+1),\quad\Re s<0
\eea
 However, the right-hand side of \eqref{repr-s} is well defined for all $s \neq 0$ and $s \neq n$, here $n\in {\mathbb Z}_+$ and we denote it by $\cI (s)$,
\bea
\label{cI}
\cI(s)= -\,\Gamma (-s)\,\zeta(-s+1).\eea
The function $\cI(s)$ given by \eqref{cI} is a meromorphic function for $s\in {\mathbb C}$. It has poles at $s=n>0$ and a double pole at $n=0$. We define $\cI_{ren}(n)$ as
\bea
\label{cIn}
\cI_{ren}(n)&\equiv&\lim _{\epsilon \to 0}\left[-\Gamma(-n+\epsilon)\zeta(1-n+\epsilon)
 -\mbox{Pole Part}\left[ (-\Gamma(-n+\epsilon)\zeta(1-n+\epsilon)\right]\right]\nn\\&\,&
\mbox{ at point}\quad  n=1,2,3,...\label{cIn-m}\eea
and 
\bea
\label{cI0}
\cI_{ren}(0)&\equiv&\lim _{\epsilon \to 0}\left[-\Gamma(\epsilon)\zeta(1+\epsilon)
 -\mbox{Double Pole Part}\left[ (-\Gamma(\epsilon)\zeta(1+\epsilon)\right]\right]
 \\\,\nn\\\label{cIs}
 \cI_{ren}(s)&\equiv&\cI,\quad s>0, s\neq {\mathbb Z}_+ \,.
 \eea

\item In what follows we  prove that
\bea\label{IcIn}
 \cI_{ren}(n)
&=& I_{ren}(n),\\\label{IcIs}
 \cI(s)
&=& I_{ren}(s), \quad s\neq n
\eea
The detail definitions of $I_{ren}(n)$ and 
$\cI_{ren}(n)$ will be given in Sect.\ref{CutR} and Sect.\ref{AnalR}, respectively. In Sect. \ref{Equiv} we show the equivalence of these two forms of renormalizations, i.e. validity of \eqref{IcIn} and 
\eqref{IcIs}.
\end{itemize}
\newpage
\section{Cut-off renormalization }\label{CutR}
In this section we present the explicit form of the renormalized version of \eqref{Isa}  after the minimal renormalization.
We distinguish two cases: integer and non-integer $s\geq0$.
\begin{itemize}

\item For $s=n$,  $n=0,1,2,...$, the following proposition holds. 
\end{itemize}

{\bf Proposition 1.} {\it The renormalized version of \eqref{Isa} after minimal renormalizations  defined by \eqref{Iren} is given by}
\bea\nn
I_{ren}(n)&=&
\int _0^1\frac{1}{x^{n+1}}\Big[\ln\Big(\frac{1-e^{- x}}{x}\Big)-\sum _{k=1}^{n}c_kx^k\Big]\,dx\\\label{TFP}
&-&\frac{1}{n^2}+
\sum _{k=1}^{n-1} \frac{c_k}{k-n}
+\int _1^\infty\frac{1}{x^{n+1}}\ln\Big(1-e^{- x}\Big)dx,\quad n>0;\\\,\nn\\\label{TFP0}
I_{ren}(0)\,&=&\int _0^1\frac{1}{x}\ln\Big(\frac{1-e^{- x}}{x}\Big)\,dx+\int _1^\infty\frac{1}{x}\ln\Big(1-e^{- x}\Big)dx.
\eea

\begin{itemize}
\item For $s\neq 0$, $s\neq n\in {\mathbb Z}_+$,   in the following proposition holds.
\end{itemize}

{\bf Proposition 1$\,^\prime$.} {\it The renormalized version of \eqref{Isa} after minimal renormalizations is}
\bea\nn
I_{ren}(s)&=&\int _0^1\frac{1}{x^{s+1}}\Big[\ln\Big(\frac{1-e^{- x}}{x}\Big)-\sum _{k=1}^{n(s)}c_kx^k\Big]\,dx\\\label{TFP-m}
&-&\frac{1}{s^2}+
\sum _{k=1}^{n(s)} \frac{c_k}{k-s}
+\int _1^\infty\frac{1}{x^{s+1}}\ln\Big(1-e^{- x}\Big)dx
\,,\\
n(s)&= &\mbox{Entier[$s$], i.e the integer part of } \, s. 
\eea

{\bf Remark.} The formula \eqref{TFP0} can be considered as a generalization of the Chebyshev formula for the zeta-function, see \cite{Cheb,Bombieri}. 
\\

To prove these propositions we present $I(s,a)$ given by \eqref{Isa}
as 
\bea\label{decom-1}
I(s,a)&=&
I(s,a,1)+I(s,1,\infty),
\eea
where
\bea\label{decom-a}
I(s,a,1)&=&\int _{a}^{1}\ln
\Big(1-e^{- x}\Big)\, \frac{dx}{x^{1+s}}, \qquad a<1\\
\label{decom-b}
I(s,1,\infty)&=&
\int _{1}^{\infty}\ln
\Big(1-e^{- x}\Big)\, \frac{dx}{x^{1+s}}.
\eea
We expand the  integrand in \eqref{decom-a} in the power series near the $x=0$. We have
\bea\label{log-exp}
\ln\Big(1-e^{- x}\Big)=\log(x)+\sum _{k=1}^{\infty}c_k x^k,\eea
$c_k$ are related with the Bernoulli numbers $B_k$, see Appendix A,
 \be
c_k=
\frac{1}{k\,k!}\,B_{k}\label{ckB}\ee
and  we have
\bea\label{exp}
\frac{1}{x^{1+s}}\ln\Big(1-e^{- x}\Big)=\frac{1}{x^{1+s}}\log(x)+\sum _{k=1}^{n(s)}c_k x^{k-1-s}+\sum _{k=n(s)+1}^{\infty}c_k x^{k-1-s}
\eea
We take $n(s)=E[s]$, where $E[s]$ is the integer part of $s$. Therefore, in the first sum in the RHS of \eqref{exp} all terms have power less then $-1$ and after integrating 
 the equality \eqref{exp} in interval $(a,1)$
give raise to singular terms for $a\to 0$. Let us find these singular terms explicitly first for $s=n.$

\begin{itemize}
\item $s=n$. We have
\end{itemize}
\bea
I(n,a,1)&=&\int_a^1\Big[\ln\left(\frac{1-e^{- x}}{x}\right)-\sum _{k=1}^{n}c_k x^k\Big]\frac{dx}{x^{1+n}}+
\int _a ^1\frac{\log x}{x^{1+n}}dx+\sum _{k=1}^{n}c_k\int _a ^1 \frac{dx}{x^{1+n-k}}\nn\\\label{pfp1}
&=&\int_a^1\frac{1}{x^{1+n}}\left[\ln\left(\frac{1-e^{- x}}{x}\right)-\sum _{k=1}^{n}c_k x^k\right]dx\\\label{pfp}
&+&\frac{1}{n^2a^{n}}+\frac{ \log a}{na^{n}}-c_n \log a-\frac{1}{n^2}
+\sum _{k=1}^{n-1}c_k\left[\frac{1}{k-n}-\frac{a^{k-n}}{k-n}\right]
\eea
This identity  gives the representation
\bea\label{FRn}
I(n,a,1)=
S(n,a)+F(n) +\cO(a),\eea
where $S(a,n)$ includes all singular terms at $a\to 0$
\bea
\label{Sn}
S(n,a)=\frac{ \log a}{na^{n}}+\frac{1}{n^2a^{n}}-c_n \log a-\sum _{k=1}^{n-1}c_k\frac{a^{k-n}}{k-n},\eea
 $F(n)$ is the finite part that contains the limit at $a\to 0$ of the convergent integral 
in the  line \eqref{pfp1} and two terms from the line  \eqref{pfp}
\bea
-\frac{1}{n^2}+\sum _{k=1}^{n-1} \frac{c_k}{k-n}
\eea
The representation \eqref{FRn} gives the statement of Proposition 1.
\begin{itemize}
\item For arbitrary  $s>0$ and $s=n+\delta,$ $0<\delta<1$ we have
\end{itemize}
\bea\nn
I(s,a,1)&=&\int_a^1\Big[\ln\left(\frac{1-e^{- x}}{x}\right)-\sum _{k=1}^{n(s)}c_k x^k\Big]\frac{dx}{x^{1+s}}+
\int _a ^1\frac{\log x}{x^{1+s}}dx+\sum _{k=1}^{n(s)}c_k\int _a ^1 \frac{dx}{x^{1+s-k}}\nn\\\label{pfpm}
&!!!=&\int_a^1\Big[\ln\left(\frac{1-e^{- x}}{x}\right)-\sum _{k=1}^{n(s)}c_k x^k\Big]\frac{dx}{x^{1+s}}\\\label{SS}
&+&\frac{1}{s^2a^{s}}+\frac{ \log (a)
   }{sa^{s}}-\frac{1}{s^2}
+\sum _{k=1}^{n(s)}c_k\left[\frac{1}{k-s}-\frac{a^{k-s}}{k-s}\right],
\eea
$n(s)$ is the integer part of $s$. This identity gives representation
\bea
I(s,a,1)=
S(s,a)+F(s) +\cO(a),\eea
where $S(s,a)$ includes all singular terms at $a\to 0$
\bea
S(s,a)=\frac{1}{(s)^2a^{s}}+\frac{ \log (a)}{sa^{s}}
+\sum _{k=1}^{n(s)}c_k\left[-\frac{a^{k-n-\delta}}{k-n-\delta}\right].\eea
Few terms give contributions to $F(s)$. The integral in the line \eqref{pfpm} converges at $a\to 0$ and contributes to the finite part $F(s)$. Two  terms 
\bea
\label{Fs}
-\frac{1}{s^2}+\sum _{k=1}^{n(s)}\frac{1}{k-s}c_k\eea
also contribute to the final part $F(s)$ and we get
\bea
F(s)&=&\int_0^1\Big[\ln\left(\frac{1-e^{- x}}{x}\right)-\sum _{k=1}^{n(s)}c_k x^k\Big]\frac{dx}{x^{1+s}}-\frac{1}{s^2}+\sum _{k=1}^{n(s)}\frac{1}{k-s}c_k\eea
Subtracting $S(s,a)$ and removing regularization we get  the proof of the Proposition $1^\prime$.
$$\,$$

\section{Analytical renormalization}\label{AnalR}
In this section we present the explicit form of the renormalized version of \eqref{Isa}  after the analytical  renormalization. As in Sect.\ref{CutR}
we distinguish two cases: integer and non-integer $s\geq0$.
\begin{itemize}

\item For $s=n$,  $n=0,1,2,...$ the following proposition holds.
\end{itemize}

{\bf Proposition 2.} {\it The renormalized version of \eqref{repr-s} after analytical renormalizations defined by \eqref{cIn} is given by}

\bea\label{dc}
\cI_{ren}(n)=-\left\{
\begin{array}{cc}
    \frac{(-1)^n}{n!}\,\left[  \zeta '(1-n)+\left(-\gamma+\sum_{k=1}^{n}\frac{1}{k}\right)\,\zeta (1-n)\right],  &  n=1,2,3...\\\,\\
    \,\frac{1}{12} \left(12 \gamma
   _1+6 \gamma ^2-\pi
   ^2\right),  & n=0
\end{array}\right.
\eea

To prove this Proposition we follow the definitions \eqref{cIn} and   take  $s=n-\epsilon$, $n\neq 0$ and \eqref{cI} for $n=0$. We have
\bea\nn
&\,&\Gamma (-s)\,\zeta(-s+1)=\Gamma(\epsilon-n)\zeta(1-n+\epsilon)\\
&=&\frac{(-1)^n}{n!}\,\zeta (1-n)\, \frac{1}{\epsilon}
+\frac{(-1)^n}{n!}\,\left[  \zeta '(1-n)+\left(-\gamma+\sum_{k=1}^{n}\frac{1}{k}\right)\,\zeta (1-n)\right]+\cO(\epsilon)\label{ARn}\eea
 and for  $n=0$  we have 
   \bea\label{AR0}
   -\Gamma (-\epsilon ) \zeta
   (1-\epsilon )&=&\frac{1}{12} \left(12 \gamma
   _1+6 \gamma ^2-\pi
   ^2\right)-\frac{1}{\epsilon ^2}+\cO(\epsilon),\eea
   where $\gamma$  is the Euler–Mascheroni constant, $\gamma=0.577$ and $\gamma_1$ is the Stieltjes constant, $\gamma_1=-0.0728$.
   Subtracting the pole in \eqref{ARn} and double pole in \eqref{AR0} we get the first line and the second line in \eqref{dc}, respectively.
   
   {\bf Proposition $2^\prime$.} {\it The analytical regularization  for $s\neq {\mathbb Z}$ gives directly the finite   answer $\cI(s)$}.
   
   The proof follows immediately from the form of $\cI(s)$ given by \eqref{cI}.\\
   
   {\bf Remark.} Note that we have considered here  the integral \eqref{I(s)} as a whole. However, it should be noted that this integral \eqref{I(s)}  is equal to the product of the gamma function and the zeta function  and in fact the divergences occur only in the gamma function.  In this case, it is possible to carry the gamma function renormalization and obtain similar results.

In this case instead of the expression \eqref{cIren-n}
we get
\bea
\label{IG}
\cI_{ren,\Gamma}(n)=\frac{(-1)^n}{n!}\, \left(-\gamma+\sum_{k=1}^{n}\frac{1}{k}\right)\,
\zeta (1-n)\eea
 By using  \eqref{AB:B1} we get

 \bea\label{IGm}
\cI_{ren,\Gamma}(n)&=&
% \frac{(-1)^n}{n!}\, \left(-\gamma+\sum_{k=1}^{n}\frac{1}{k}\right)\,
% \frac{(-1)^{n-1} B_{n}}{n}\\&=&
\frac{ B_{n}}{n!\,n}\left(\gamma-\sum_{k=1}^{n}\frac{1}{k}\right)
\eea

\section{Equivalence of cut-off and analytical renormalizations}\label{Equiv}
In this Section we prove that the renormalized free energies defined by  the cut-of  renormalization \eqref{Iren} and the analytical regularization \eqref{cIn}-\eqref{cIs} coincide. We distinguish three  cases: $s=n\neq 0$, $s=0$ and $s\neq 0,n\in {\mathbb Z}_+$.
\\

{\bf Proposition 3}. {\it The minimal renormalized free energy \eqref{Iren} for  $s=n\neq 0$ and the analytic renormalized free energy \eqref{cIn} coincide}

\be I_{ren}(n)=\cI_{ren}(n).\label{IcI=}\ee
{\it Explicitly \eqref{IcI=} means the validity of the following identity} 
\bea\nn
&&
\int _1^\infty\frac{1}{x^{n+1}}\ln\Big(1-e^{- x}\Big)dx
+
\int _0^1\frac{1}{x^{n+1}}\Big[\ln\Big(\frac{1-e^{- x}}{x}\Big)-\sum _{k=1}^{n}c_kx^k\Big]\,dx
\\\nn
&&\qquad-\frac{1}{n^2}+
\sum _{k=1}^{n-1} \frac{c_k}{k-n}\\\nn\\
\label{Eq}
&=&\frac{(-1)^n}{n!}\,\left[  \zeta '(1-n)+\left(-\gamma+\sum_{k=1}^{n}\frac{1}{k}\right)\,\zeta (1-n)\right],\quad n=1,2,...,
\eea
{\it for}
\bea
c_k&=&-\frac{(-1)^k}{k!}\,\zeta (1-k)=\frac{B_{k}}{k\,k!},\qquad k=1,2,3,...\,.\label{ckk}
\eea
$\,$

{\bf Proof.}
Let us consider the function $\psi (n,s)$
of $s$-variable depending on the integer parameter $n$, $n>0$ , defined for $\Re s< n+1$ as
\bea\nn
\psi (n,s)&=&-\frac{1}{ s^2}+
\sum _{k=1}^{n-1} \frac{c_k}{k-s}
+\int _1^\infty\frac{1}{x^{s+1}}\ln\Big(1-e^{- x}\Big)dx\\\label{Psis}
&+&
\int _0^1\frac{1}{x^{s+1}}\Big[\ln\Big(\frac{1-e^{- x}}{x}\Big)-\sum _{k=1}^{n}c_kx^k\Big]\,dx.
\eea
According Proposition 1,
\bea\label{PsiIr}
\psi (n,n)&=&I_{ren}(n).\eea
For $s<0$  the integral 
\be \int _0^1\frac{1}{x^{s+1}}\ln\Big(1-e^{- x}\Big)\,dx
\ee
converges and after rearrangement of the terms in the RHS of \eqref{Psis} we can rewrite $\psi (n,s)$ as
\bea\label{PHTG}
\psi (n,s)&=&
H(n,s)-\,\Gamma (-s)\,\zeta(-s+1)-T(n,s), \quad s<0,
\eea
where 
\bea\nn
H (n,s)&=&-\frac{1}{s^2}+\sum _{k=1}^{n-1} \frac{c_k}{k-s},\\
T (n,s)&=&\int _0^1\frac{1}{x^{s+1}}\Big[\ln x+\sum _{k=1}^{n}c_kx^k\Big]\,dx.
\eea
Evaluating $T(n,s)$  for  $\Re s<0$  we get 
\bea
T(n,s)=-\frac{1}{s^2}+\sum _{k}^{n}\frac{c_k}{k-s}\eea
 and the RHS of \eqref{Psis} becomes equal to
\bea
\label{GZR}
-\,\Gamma (-s)\,\zeta(-s+1)-\frac{c_n}{n-s},\eea
that is  meromorphic function of variable $s$
on whole $\mathbb C$.
 This function from one side due to equation \eqref{PsiIr} for 
$s= n$ coincides with $I_{ren}$ and from other side  can be evaluated in the following way. 

First note that the pole in \eqref{GZR} is exactly the  pole  that has to be subtracted in the analytical renormalization defined in \eqref{cIn}. For this purpose we take  $s=n-\epsilon$
and consider $\Gamma (-s)\,\zeta(-s+1)$
for small $\epsilon$. Due to \eqref{App:GZ}, see Appendix \ref{Appen:B}
we have 
\bea\nn
&\,&\Gamma (-s)\,\zeta(-s+1)=\Gamma(\epsilon-n)\zeta(1-n+\epsilon)\\\label{PoleGn}
&=&\frac{(-1)^n}{n!}\,\zeta (1-n)\, \frac{1}{\epsilon}
+\frac{(-1)^n}{n!}\,\left[  \zeta '(1-n)+\left(-\gamma+\sum_{k=1}^{n}\frac{1}{k}\right)\,\zeta (1-n)\right]+\cO(\epsilon)\eea
Therefore, to check that the pole in \eqref{GZR} is exactly the  pole that  we
have in \eqref{PoleGn}, we have to check that 
\bea
c_n=-\frac{(-1)^n}{n!}\,\zeta (1-n)\label{ckn}\eea
The proof of equation \eqref{ckn} follows from representation of $\zeta(-n)$ in term of the Bernoulli numbers, see \eqref{AB:B1} in Appendix \ref{Appen:B}, we have
\be\label{z-n}
 \zeta(1-k)=\frac{(-1)^{1-k}\,B_{k}}{k}.
\ee
Due to \eqref{z-n} the  RHS of \eqref{ckn} is
\be
-\frac{(-1)^n}{n!}\,\zeta (1-n)=-\frac{(-1)^n}{n!}\,\frac{(-1)^{1-n}\,B_{n}}{n}=
\frac{1}{n\,n!}\,B_{n}\label{ckBm}\ee
and the obtained expression coincides with definition \eqref{ckB}
of $c_k$.

Also from \eqref{PoleGn} we get
\bea
\label{cIren-n}
\cI_{ren}(n)=\frac{(-1)^n}{n!}\,\left[  \zeta '(1-n)+\left(-\gamma+\sum_{k=1}^{n}\frac{1}{k}\right)\,\zeta (1-n)\right]\eea

$$\,$$
{\bf Proposition $3^\prime$}. {\it The minimal renormalized free energy \eqref{TFP} and the analytic renormalized free energy \eqref{cIs} coincide,} 

\be I_{ren}(s)=\cI(s),\quad\mbox{for}\quad s> 0\quad\mbox{and}\quad s\neq n\in {\mathbb Z}_+.\label{IcIs=}\ee
{\it Explicitly \eqref{IcIs=} means the validity of the following identity}
\bea\nn
&&
\int _0^1\frac{1}{x^{s+1}}\Big[\ln\Big(\frac{1-e^{- x}}{x}\Big)-\sum _{k=1}^{n(s)}c_kx^k\Big]
\,dx+\int _1^\infty\frac{1}{x^{s+1}}\ln\Big(1-e^{- x}\Big)dx\\\nn
&&\qquad-\frac{1}{s^2}+
\sum _{k=1}^{n(s)} \frac{c_k}{k-s}\\\nn\\
\label{Eqs}
&=&-\,\Gamma (-s)\,\zeta(-s+1),\quad n(s)=E(s) - \mbox{the integer  part of number}\,s.
\eea
To prove the identity \eqref{Eqs} we consider the function $\psi (n,s)$, $s<n+1$,
\bea\nn
\psi (n,s)&=&-\frac{1}{s^2}+
\sum _{k=1}^{n} \frac{c_k}{k-s}
+\int _1^\infty\frac{1}{x^{s+1}}\ln\Big(1-e^{- x}\Big)dx\\\label{psi}
&+&
\int _0^1\frac{1}{x^{s+1}}\Big[\ln\Big(\frac{1-e^{- x}}{x}\Big)-\sum _{k=1}^{n}c_kx^k\Big]\,dx.
\eea

From the Proposition $1^\prime$ we see  that
\bea
\psi (n(s),s)=I_{ren}(s).\eea 
From other site, for $\psi (n,s)$ at $\Re s<0$
we can write the representation
\bea\label{HTGG}
\psi (n,s)&=&
-\,\Gamma (-s)\,\zeta(-s+1), \quad \Re s<0.
\eea
Indeed, for $s<0$ we rearrange the terms in  \eqref{psi} and get 
\bea\label{HTGmm}
\psi (n,s)&=&
H(s,n)-\,\Gamma (-s)\,\zeta(-s+1)-T(n,s),
\eea
where
\bea\nn
H (n,s)&=&-\frac{1}{s^2}+\sum _{k=1}^{n} \frac{c_k}{k-s},\\
T (n,s)&=&\int _0^1\frac{1}{x^{s+1}}\Big[\ln x+\sum _{k=1}^{n}c_kx^k\Big]\,dx.
\eea
Evaluating $T(n,s)$  for  $\Re s<0$  we get \bea
T(n,s)=-\frac{1}{s^2}+\sum _{k}^{n}\frac{c_k}{k-s}\eea
and $T(n,s)$ compensates $H(n,s)$ and we 
get \eqref{HTGG}. From the uniqueness of analytical continuation we get \eqref{Eqs}.

\section{Examples}\label{Examp}
In this Section we consider few examples of specific values of $d,\,D,\,\alpha$ which provide the Bose gas interpretations of the Schwarzschild black hole thermodynamics.
For any $D=4,5,6,..$ and $d=1,2,3,...$,
we set $\alpha =d/(2-D)$. Using \eqref{fe-d-nx2} 
we get
\bea\label{fe-d-nxr}F_{BG,ren}&=&\frac{\Omega_{d-1}}{ \alpha}\left(\frac{L}{2\pi}\right)^d\, \lambda^{D-2}\,I_{ren}(D-2)\,\beta^{D-3}.
\eea

Considering the equation \eqref{App:IRn0}-\eqref{App:IRn4} we obtain the following expressions for the Bose gas free energy
\begin{itemize}
\item $-\frac{d}{\alpha}=2$.  In this case  $D=4$  and according \eqref{App:IRn2}
\bea\label{App:IRn2}
\cI_{ren}(2)=\frac{1}{48} (24 \log (A)+1-2 \gamma )=0.121,
\eea
therefore,
\bea
F_{BG,ren}&=&-\frac{2\Omega_{d-1}}{d }\left(\frac{L}{2\pi}\right)^d\, \lambda^2\,I_{ren}(2)\beta=
-0.242\,\frac{\Omega_{d-1}}{d }\left(\frac{L}{2\pi}\right)^d\, \lambda^2\,\beta.
\eea
This case is not suitable for us since it gives negative entropy.
\item $-\frac{d}{\alpha}=3$.  In this case  $D=5$  and according \eqref{App:IRn3}
\bea\label{App:IRn3}
\cI_{ren}(3)=\frac{1}{6} \zeta '(-2)=-0.00507,\eea
therefore we have
\bea\label{App:FBGD5}
F_{BG,ren}&=&-\frac{3\Omega_{d-1}}{ d}\left(\frac{L}{2\pi}\right)^d\, \lambda^{3}\,I_{ren}(3)\,\beta^{2}=
0.0152\,\frac{\Omega_{d-1}}{ d}\left(\frac{L}{2\pi}\right)^d\, \lambda^{3}\,\beta^{2}.
\eea
\item $-\frac{d}{\alpha}=4$.  In this case  $D=6$  and according \eqref{App:IRn4}
\bea\label{App:IRn4}
\cI_{ren}(4)=\frac{-1440 \zeta '(-3)-25+12 \gamma }{34560}=-0.000747,\eea
therefore we have
\bea
\label{App:FBGD6}
F_{BG,ren}&=&-\frac{4\,\Omega_{d-1}}{ d}\left(\frac{L}{2\pi}\right)^d\, \lambda^{4}\,I_{ren}(4)\,\beta^{3}=0.00299\,
\frac{\Omega_{d-1}}{ d}\left(\frac{L}{2\pi}\right)^d\, \lambda^{4}\,\beta^{3}.
\eea
\end{itemize}

From the consideration above, we see that among the listed cases, only in the cases
$D=5,6$ we obtain a positive value of the corresponding entropy.
\begin{itemize}
\item $D=5$.  In this case   according \eqref{App:FBGD5}
we have
\bea
S_{BG,ren}&=&
0.0304\,\frac{\Omega_{d-1}}{ d}\left(\frac{L}{2\pi}\right)^d\, \lambda^{3}\,\beta^{3}.
\eea
Here $d=1,2,3,...$ and the corresponding $\alpha$ takes values $-1/3,-2/3,-1,...$.
\item $D=6$.  In this case   according \eqref{App:FBGD6}
we have
\bea S_{BG,ren}&=&0.00896\,
\frac{\Omega_{d-1}}{ d}\left(\frac{L}{2\pi}\right)^d\, \lambda^{4}\,\beta^{4}.
\eea
Here $d=1,2,3,...$ and the corresponding $\alpha$ takes values $-1/4,-1/2,-3/4,...$.
\end{itemize}
$$\,$$

Let us note, that in the case of renormalization \eqref{IGm} we get
\bea\label{fe-d-nxr-G}F_{BG,ren,G}(D)&=&\frac{\Omega_{d-1}}{ \alpha}\left(\frac{L}{2\pi}\right)^d\, \lambda^{D-2}\,I_{ren,G}(D-2)\,\beta^{D-3}.
\eea

Since $\alpha<0$ the sign of \eqref{fe-d-nxr-G} is opposite to the sign of $I_{ren,G}(D-2)$ and according to 
\eqref{IGm} the sign of  $F_{BG,ren,G}$
is defined by the Bernoulli number, i.e. 
$F_{BG,ren,G}(D)<0$ for $D=4k$ and 
$F_{BG,ren,G}(D)>0$ for $D=4k+2$, $k=1,2,3$. For odd dimensions $F_{BG,ren,G}(D)=0$.

\section{Conclusion}\label{Concl}
In this paper the Schwarzschild black hole thermodynamics is modeled by the Bose gas statistical system. It is shown that the  Schwarzschild black hole in $D=5$ and $D=6$ space-time dimensions correspond to the Bose gas with one-particle energy 
$\varepsilon(k)=\lambda \, k^\alpha $  in $d$ dimensional space with $d/\alpha=-3$ and  
$d/\alpha=-4$, respectively. Divergences in these Bose gas models are discussed. It is shown that  the cut-off minimal subtraction renormalization scheme is equivalent to the analytical renormalization.This method does not work for the case of $D=4$
Schwarzschild black hole, which corresponds to the Bose gas in $d=-4$ negative dimension as it has been shown in the previous paper \cite{Arefeva:2023kpu}. The microscopic statistical mechanics description of the Schwarzschild black hole thermodynamics suggested in this and the previous papers use negative dimensions or renormalizations of the Bose gas models. \\

It would be interesting to obtain a similar microscopic description of more general black holes including the Reissner-Nordstrom, Kerr and other black holes. These models also violate the third law of  thermodynamics,  so it is natural to expect that the corresponding statistical mechanic models will also have unusual properties.

\newpage
\section*{Acknowlegments}

We would like to thank D. Ageev, V. Berezin, V. Frolov, M. Khramtsov, K. Rannu, P. Slepov, A. Teretenkov, A. Trushechkin and V. Zagrebnov for fruitful discussions. This work is supported by  the Russian Science Foundation (project
19-11-00320, V.A. Steklov Mathematical Institute).

$$\,$$
{\Large\bf Appendix.}
\appendix
\section{Bernoulli numbers and $c_k$}\label{Appen:A}
Differentiating \eqref{log-exp} one has
\be\label{Dlog-exp}
\frac{1}{ e^{x}-1}=\sum _{k=0}  k c_k\,x^{k-1}.\ee
Comparing \eqref{Dlog-exp} with the generation function for the Bernoulli numbers 
\be\label{gen-B}
\frac{x}{e^x-1}=\sum B_k \frac{x^{k}}{k!},\ee
we see that
\be
\label{A:3}
 k c_k= \frac{B_k}{k!}.\ee
that gives \eqref{ckB}.

\section{Values of $\zeta$ and $\Gamma$ functions}
Here we present some known facts about gamma and zeta functions \cite{KV}.
One has \label{Appen:B}
\bea\label{AB:B1}\zeta(-n)=\frac{(-1)^n B_{n+1}}{n+1}, \quad n=1,2,3,...
\eea
where $B_n$ are the Bernoulli numbers defined by the generating function
\eqref{gen-B}.

\be \zeta(-1)=-\frac{1}{12}; \qquad
\zeta^\prime(-1)=\frac{1}{12}-\ln A.
\ee

For $n\in\mathbb{N}$: 
\bea 
\zeta'(-2n)&=&(-1)^n\frac{(2n)!}{2(2\pi)^{2n}}\;\zeta(2n+1)
\\
-\zeta'(1-2n)&=&\left.\left(2(2\pi)^{-s}\Gamma(s)\,\zeta(s)\right)^{\,'}\right|_{s=2n}\,\cos\left(\pi\,n\right)\\
\zeta'(1-2n)&=&(-1)^{n+1}\frac{2\,\Gamma(2n)}{(2\pi)^{2n}}\Big[\left(-\log(2\pi)+\psi(2n)\right)\zeta(2n)+\zeta'(2n)\Big],\nn
\eea
here $\psi$ is  the digamma function
$$\psi(s)=\frac{\Gamma^\prime(s)}{\Gamma(s)}.$$

For $\Gamma$-function we have
\bea
\label{B:5}\frac{\Gamma(z-n)}{\Gamma(1+z)}=\frac{(-1)^n}{n!}\left(\frac{1}{z}+\sum_{r=0}^{\infty}A_r z^r\right),\\
\label{B:6}A_r=\sum_{k=1}^{n}\binom{n}{k}\frac{(-1)^{k-1}}{k^{r+1}}, \qquad
A_0=\sum_{k=1}^{n}\frac{1}{k}.
\eea
Therefore, we have
\bea\nn
\Gamma(\epsilon-n)&=&\Gamma(1+\epsilon)\,\frac{(-1)^n}{n!}\left(\frac{1}{\epsilon}+A_0+\cO(\epsilon)\right)
\\
\label{B:7}
&=&
\frac{(-1)^n}{n!}\left(\frac{1}{\epsilon}-\gamma+\sum_{k=1}^{n}\frac{1}{k}\right)+\cO(\epsilon)\eea
and
\bea\nn
&\,&\Gamma(\epsilon-n)\zeta(1-n+\epsilon)
\\\label{App:GZ}
&=&\frac{(-1)^n}{n!}\,\zeta (1-n)\, \frac{1}{\epsilon}
+\frac{(-1)^n}{n!}\,\left[  \zeta '(1-n)+\left(-\gamma+\sum_{k=1}^{n}\frac{1}{k}\right)\,\zeta (1-n)\right]+\cO(\epsilon)\eea

{\bf Particular cases of \eqref{App:GZ}}

\bea\label{App:n0}
n=0:&\,&-\Gamma(\epsilon)\zeta(1+\epsilon)=-\frac{1}{\epsilon ^2}+\frac{1}{12} \left(12 \gamma _1+6 \gamma ^2-\pi
   ^2\right)+\cO(\epsilon)\\
   \label{App:n1}
n=1:&\,&-\Gamma(\epsilon-1)\zeta(\epsilon)=-\frac{1}{2 \epsilon }+
\frac{1}{2} (-1+\gamma -\log (2 \pi ))+\cO(\epsilon)\\
 n=2:&\,&-\Gamma(\epsilon-2)\zeta(\epsilon-1)=\frac{1}{24 \epsilon }+\frac{1}{48} (24 \log (A)+1-2 \gamma )+\cO(\epsilon)\\
n=3:&\,&-\Gamma(\epsilon-3)\zeta(\epsilon-2)=\frac{1}{6} \zeta '(-2)+\cO(\epsilon)
\\
n=4:&\,&-\Gamma(\epsilon-4)\zeta(\epsilon-3)=-\frac{1}{2880
   \epsilon }+\frac{-1440 \zeta '(-3)-25+12 \gamma }{34560}+\cO(\epsilon)
\\
n=5:&\,&-\Gamma(\epsilon-5)\zeta(\epsilon-4)=\frac{1}{120} \zeta '(-4)+\cO(\epsilon).\eea
Here
$\gamma$  is the Euler–Mascheroni constant, 
\be\label{EM}\gamma=0.577...,\ee
$\gamma_1$ is the Stieltjes constant, 
\be 
\label{St}
\gamma_1=-0.0728...\ee
and $A$ is the Glaisher  constant
\be
\label{Gl}
A=1.28...\,.
\ee

   For $\cI_{ren}$ we have
   
   \bea\label{App:IRn0}
n=0:&\,&\cI_{ren}(0)=\frac{1}{12} \left(12 \gamma _1+6 \gamma ^2-\pi
   ^2\right)=-0.728694\\
   \label{App:IRn1}
n=1:&\,&\cI_{ren}(1)=
\frac{1}{2} (-1+\gamma -\log (2 \pi ))=-1.13033\\\label{App:IRn2m}
n=2:&\,&\cI_{ren}(2)=\frac{1}{48} (24 \log (A)+1-2 \gamma )=0.12116\\\label{App:IRn3m}
n=3:&\,&\cI_{ren}(3)=\frac{1}{6} \zeta '(-2)=-0.00507474
\\\label{App:IRn4m}
n=4:&\,&\cI_{ren}(4)=\frac{-1440 \zeta '(-3)-25+12 \gamma }{34560}=-0.000747065
\\
n=5:&\,&\cI_{ren}(5)=\frac{1}{120} \zeta '(-4)=0.0000665318\eea


\begin{thebibliography}{100}
 \bibitem{BCH}
B.Bardeen, B.Carter and S.W.Hawking, The four laws of black hole mechanics, {\sl Commun. Math. 
Phys.} {\bf 31}, 161 (1973).

 \bibitem{Bekenstein:1973}
J.~D. Bekenstein, ``Black holes and entropy,''
 Phys. Rev. D
  {\bf 7} (1973) 2333--2346.
\bibitem{Arefeva:2023kpu}
I.~Aref'eva and I.~Volovich,
``Violation of the Third Law of Thermodynamics by Black Holes, Riemann Zeta Function and Bose Gas in Negative Dimensions,''
[arXiv:2304.04695 [hep-th]].
\bibitem{VSV} V. S. Vladimirov,  "Generalized functions in mathematical physics". Moscow  Izdatel Nauka, (1976).
\bibitem{LLV} L. D. Landau, E. M. Lifshitz,  Statistical Physics: Volume 5.  Elsevier (2013).
\bibitem{VZ} V.A. Zagrebnov and J.-B. Bru, "The Bogoliubov model of weakly imperfect Bose gas." Physics Reports 350.5-6 (2001) 291-434.
\bibitem{Cheb} P. L. Tchebychev, "First memoir on the distribution of prime numbers", Academy of St. Petersburg,  May 24, 1848.
\bibitem{Bombieri} E. Bombieri, "Problems of the millennium: The Riemann hypothesis". Clay Mathematics Institute (2000).
% https://www.wstein.org/edu/2007/simuw07/misc/Official_Problem_Description.pdf

\bibitem{KV} A. A. Karatsuba and S. M. Voronin. "The Riemann Zeta-Function", de Gruyter, Berlin (1992).
\end{thebibliography}
\end{document}